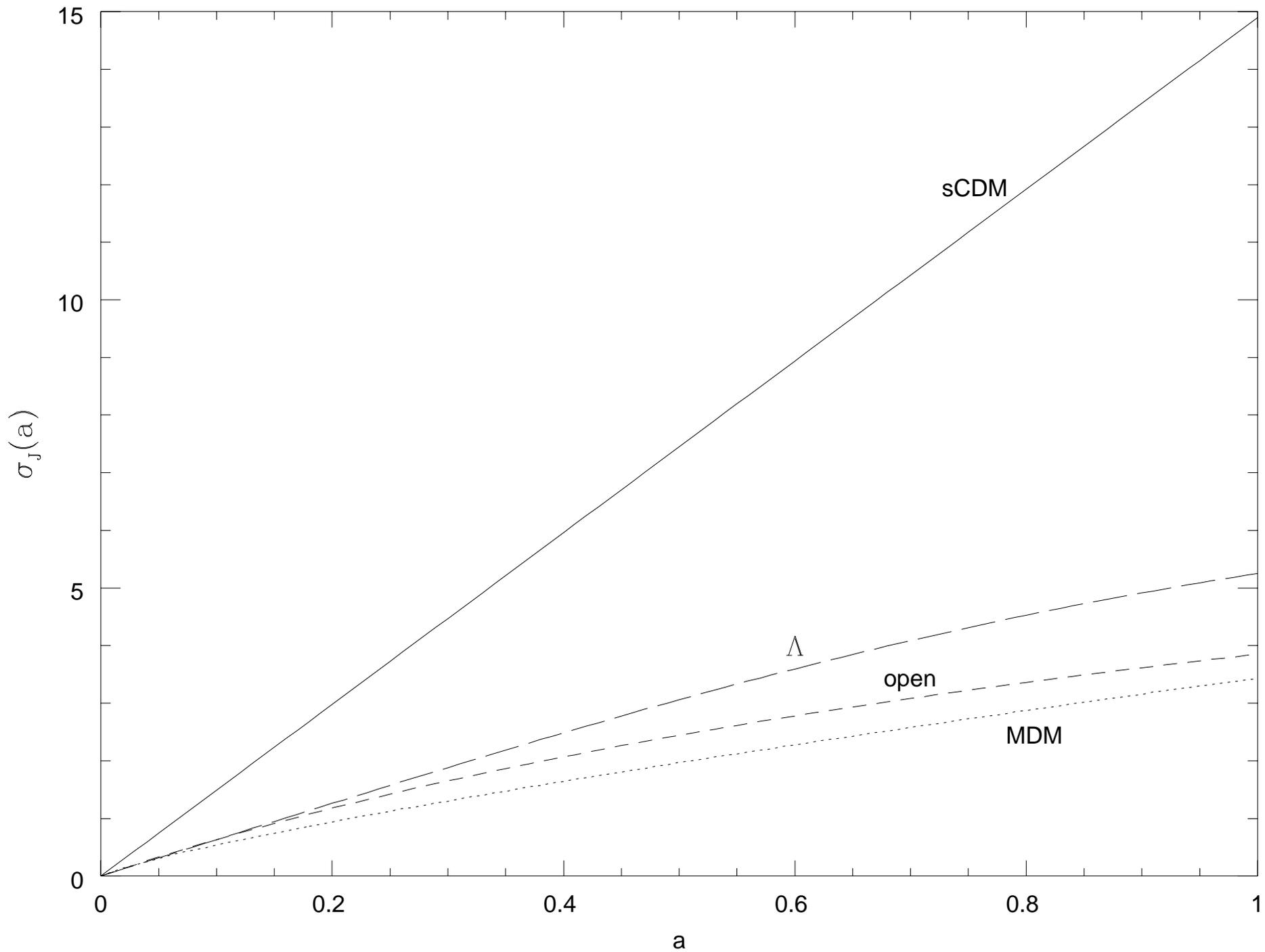

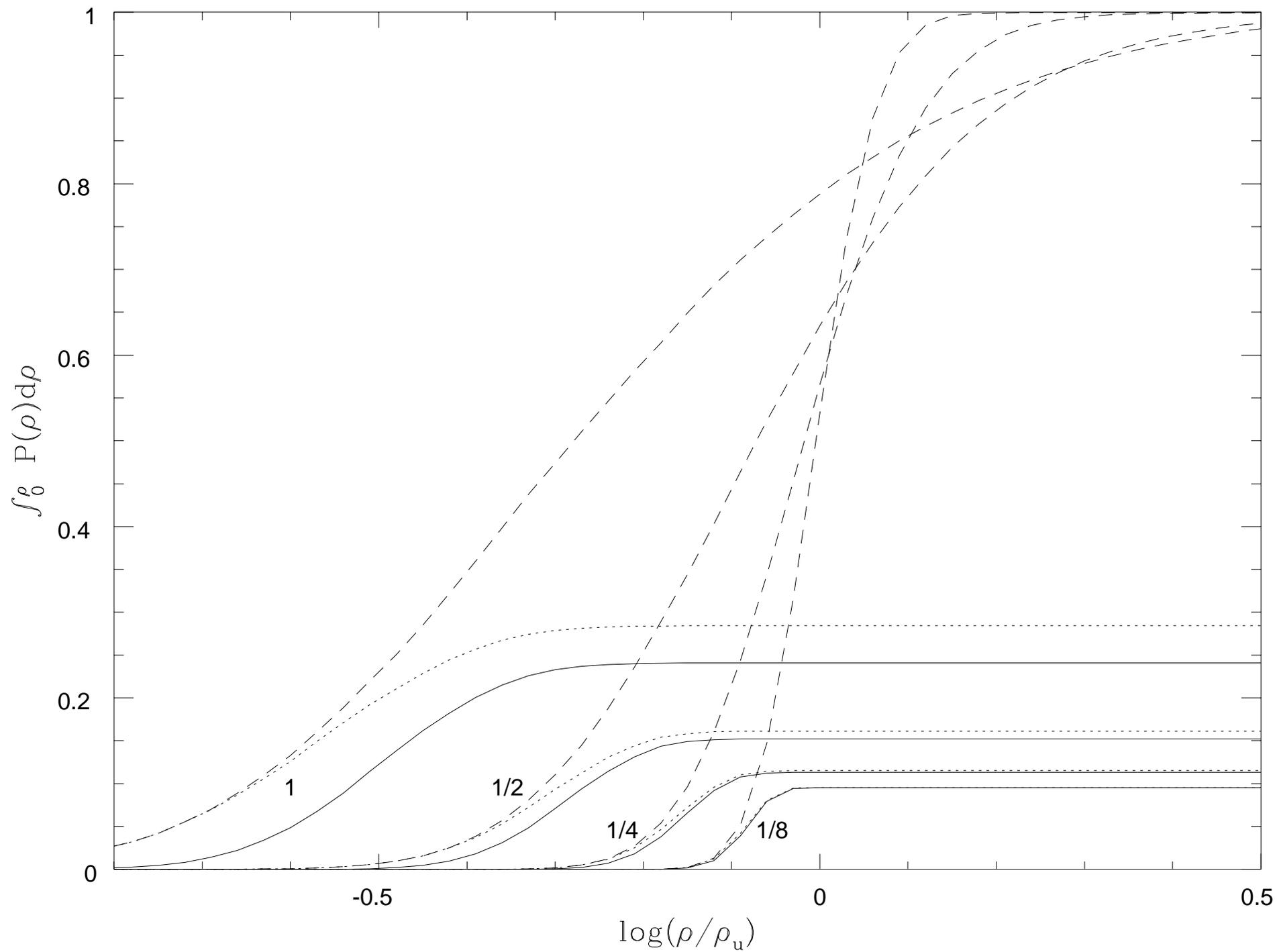

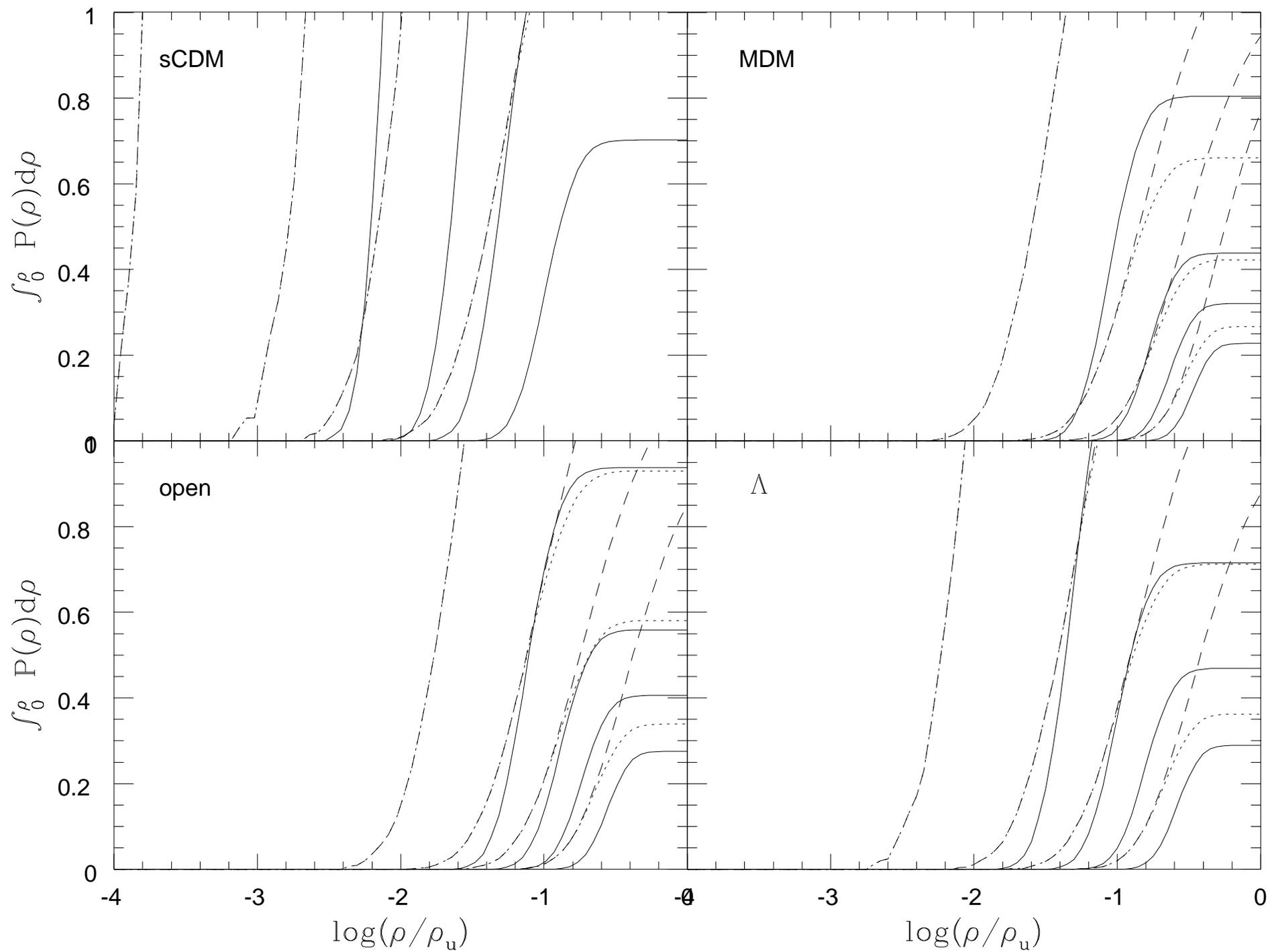

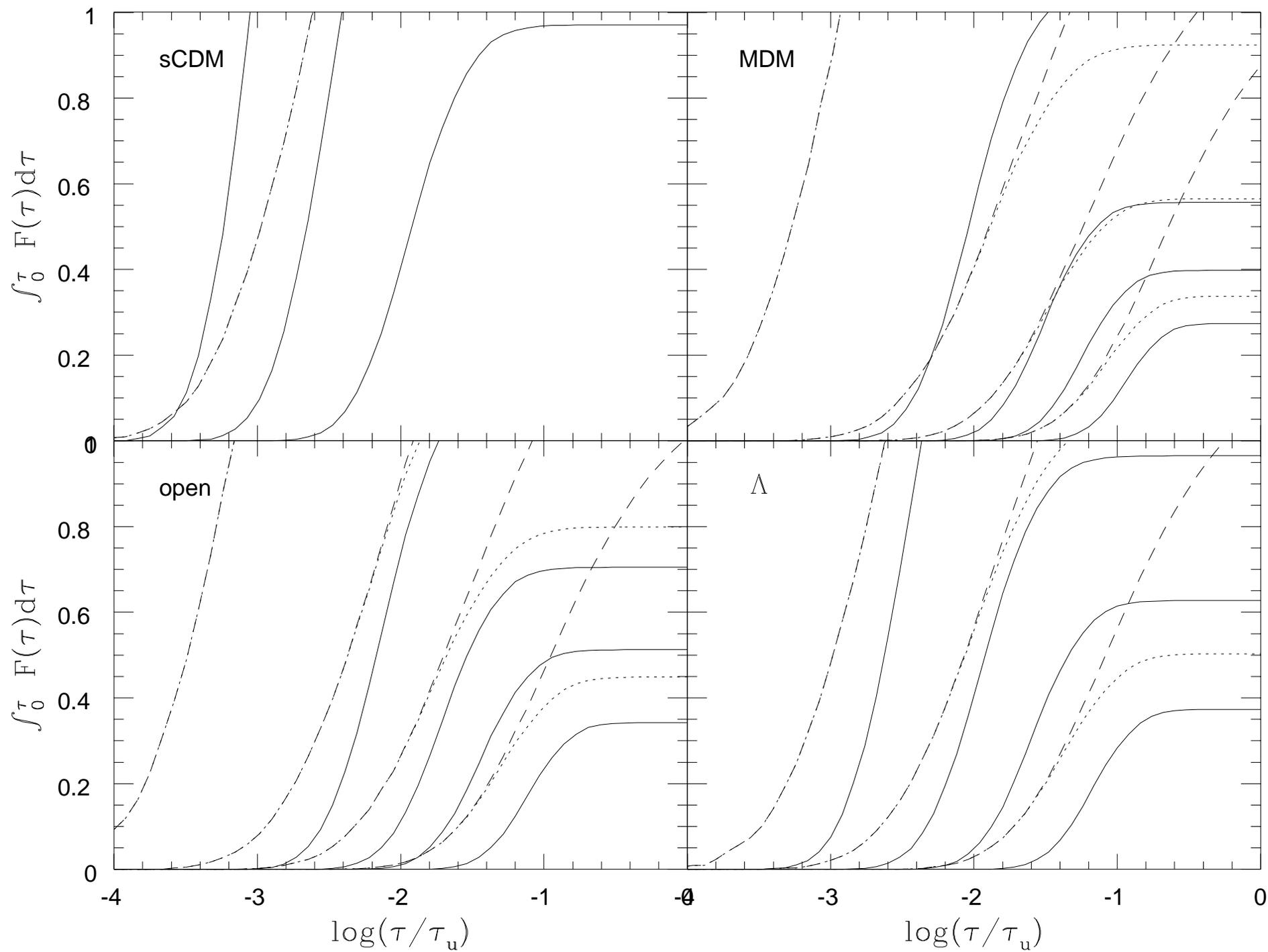

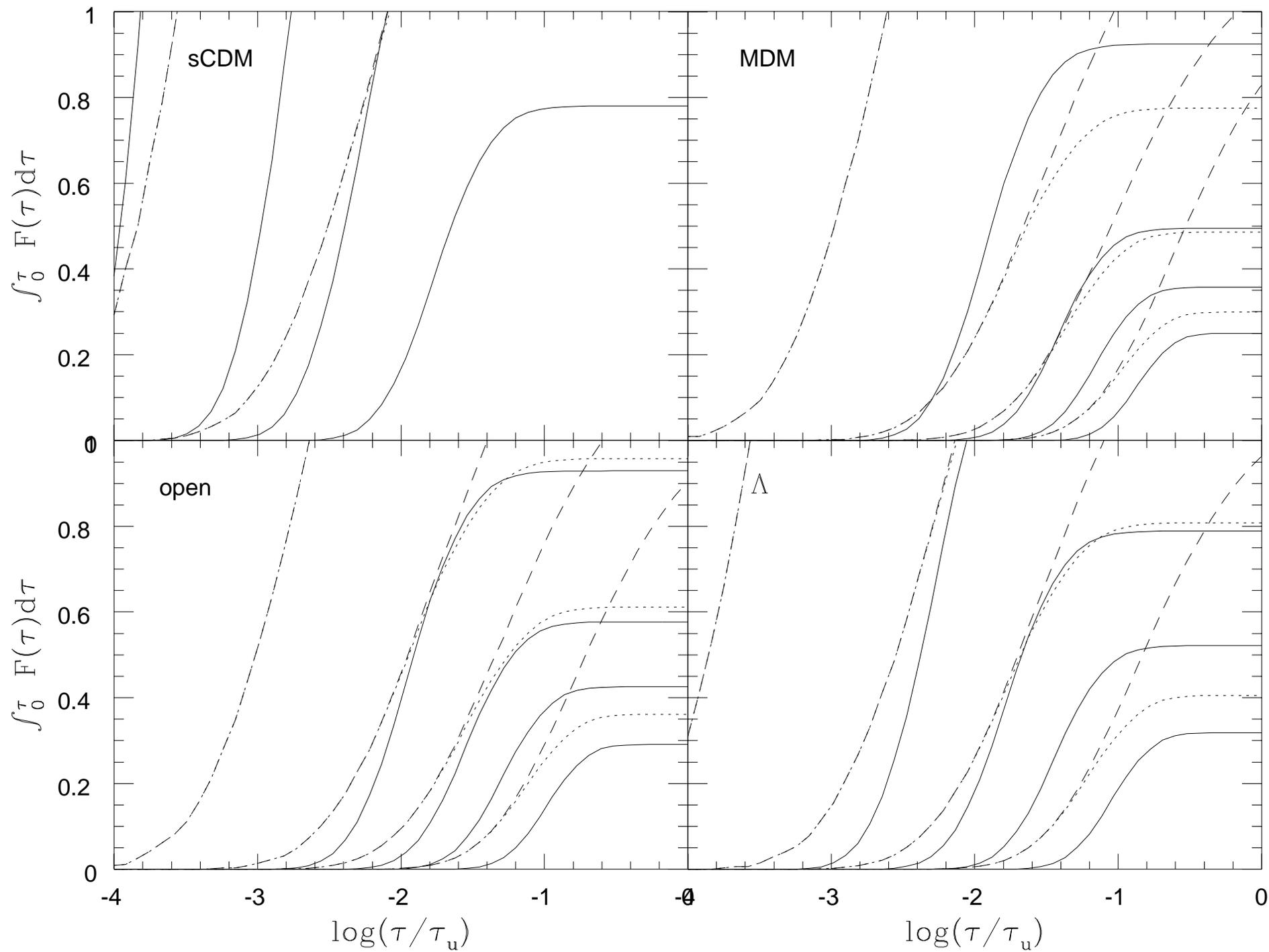

# THE GUNN-PETERSON EFFECT FROM UNDERDENSE REGIONS
## IN A PHOTOIONIZED INTERGALACTIC MEDIUM

Andreas Reisenegger[1] and Jordi Miralda-Escudé[2]

Institute for Advanced Study, Princeton, NJ 08540

## ABSTRACT

Limits to the Gunn-Peterson effect due to neutral hydrogen have generally been obtained by modeling the observed Ly$\alpha$ forest as a superposition of absorption lines with Voigt profiles arising from clouds of photoionized gas, and a hypothesized uniform continuum arising from the intergalactic medium. However, owing to the formation of structure by gravitational instability, a photoionized intergalactic medium should be inhomogeneous on scales larger than the Jeans scale, and therefore the optical depth should fluctuate. Such a fluctuating continuum can always be modeled as a superposition of lines, but this decomposition does not necessarily have a direct physical meaning.

We present a calculation of the evolution of the density in voids in a photoionized intergalactic medium, using the Zel'dovich approximation and another analytical approximation which we argue should be more accurate in this regime. From this, we calculate the probability distribution of the Gunn-Peterson optical depth in terms of the amplitude of the primordial density fluctuations. Over most wavelengths in a quasar spectrum, the optical depth originates from gas in underdense regions, or voids. Individual absorption lines should be associated with overdense regions, which we do not treat here. This causes the median Gunn-Peterson absorption to be lower than the value for a uniform medium containing all the baryons in the universe by a large factor, which increases as gravitational collapse proceeds. The Gunn-Peterson effect is the only known method to directly observe underdense matter in the universe, and it can be sensitive to the primordial fluctuations even in the non-linear regime. In particular, in the He II Gunn-Peterson effect recently detected by Jakobsen et al. , gaps in the absorption are a very sensitive probe to the most underdense voids.

We apply our calculations to the observations of the intensity distribution in a $z = 4.11$ quasar by Webb and coworkers. We show that if Ly$\alpha$ clouds arise from gravitational collapse, their observations must be interpreted as the first detection of the fluctuating Gunn-Peterson effect, with a median value $\tau_{GP} \simeq 0.06$ at $z = 4$. If the linearly extrapolated rms density fluctuation at the Jeans scale for the photoionized gas were close to unity at this redshift (which is the case in typical low-density models with cold dark matter), then $\tau_{GP}$ should be $\sim 1/5$ of the optical depth that would be produced by a uniform intergalactic medium. This is consistent with the predicted baryon density from primordial nucleosynthesis, and the intensity of the ionizing background derived from the proximity effect. From the numerical simulations of Cen et al. , such models also predict correctly the number of Ly$\alpha$ absorption lines observed. For theories with much larger density fluctuations (such as standard cold dark matter), we argue that, given the observed number of lines with $N_{HI} \gtrsim 10^{14}\,\mathrm{cm}^{-2}$, the Gunn-Peterson optical depth should be much lower than observed; this needs to be investigated in more detail using numerical simulations.

---

[1] E-mail: andreas@guinness.ias.edu

[2] E-mail: jordi@guinness.ias.edu

*Subject headings*: intergalactic medium - quasars: absorption lines - cosmology: large-scale structure of Universe

## 1. INTRODUCTION

If the baryonic material that is seen today in galaxies (and X-ray gas) had in the past been in the form of neutral gas distributed in the intergalactic medium, a trough should have been seen in the spectrum of any sufficiently distant quasar lying behind such a neutral medium at wavelengths below the Ly$\alpha$ line. The trough is caused by scattering of photons as they are redshifted through the frequency of the Ly$\alpha$ line (Gunn & Peterson 1965). The absence of the trough implies that the medium had been ionized before the epoch of the highest redshift quasar known ($z = 4.89$; Schneider, Schmidt, & Gunn 1991).

If the ionization is due to photons, an abundance of neutral gas in photoionization equilibrium should still be present. This should yield a depression of the flux below the Ly$\alpha$ line, which would be constant for homogeneously distributed neutral hydrogen. What is observed instead is the Ly$\alpha$ forest, caused by a rapidly varying optical depth to Ly$\alpha$ scattering with wavelength. This reflects a highly inhomogeneous distribution of neutral hydrogen.

Since the discovery of the Ly$\alpha$ forest, there have been several attempts to detect a uniform Gunn-Peterson effect, by subtracting the contribution to the average depression of the flux of the individually detected absorption lines. This has led to progressively more stringent lower limits to the intensity of the ionizing background, if the density of the intergalactic medium is assumed to be close to the baryonic density required by primordial nucleosynthesis (Steidel & Sargent 1987; Jenkins & Ostriker 1991; Webb et al. 1992; Giallongo et al. 1993, 1994). Since this intensity is also constrained by the observation of the proximity effect (e.g., Bajtlik, Duncan, & Ostriker 1988), this could eventually lead to the conclusion that the intergalactic medium must be collisionally ionized.

These limits are based on the assumption of the presence of a uniform intercloud medium, as distinct from individual clouds producing the absorption lines. However, there does not seem to be any theoretical justification for such a distinction for the case of a photoionized intergalactic medium. The galaxies and clusters we observe today tell us that the universe is not homogeneous. They must have collapsed from density fluctuations which were already substantial even at the highest redshifts of observed quasars. Moreover, in any theory of hierarchical gravitational collapse, fluctuations on subgalactic scales are even larger, and collapse at an earlier epoch. Since the Jeans mass for a photoionized medium (with a temperature $T \sim 10^4$ K) is $\sim 10^{10}[(1+z)/5]^{-3/2} M_\odot$, the gas down to these very small scales should have a fluctuating density, and a uniform intergalactic medium cannot exist. Therefore, *the Gunn-Peterson effect cannot be uniform.*

Primordial fluctuations on small scales should cause the photoionized gas to collapse first in objects of the Jeans mass at $T \simeq 10^4$ K. This idea led to the minihalo model (Rees 1986; Ikeuchi 1986), and other alternatives also based on gravitational collapse as the origin of the Ly$\alpha$ clouds responsible for the observed absorption lines (e.g., Bond, Szalay, & Silk 1988). Any residual Ly$\alpha$ absorption between the lines would then mostly come from underdense regions, or voids, that are left as the gas collapses gravitationally and forms structures. The optical depth in such spectral regions can be reduced by a large amount if voids are underdense by large factors. Yet, *because the pressure of the photoionized gas prevents it from collapsing on scales smaller than the Jeans scale, an intergalactic medium pervading all space must be present, and the inhomogeneous distribution of gas in voids must yield a fluctuating Gunn-Peterson effect.*

Recently, Cen et al. (1994) presented a cosmological simulation of the Lyα forest, computing the evolution of photoionized gas when structure on the relevant scales of $50 - 500h^{-1}$ kpc in comoving units collapses gravitationally. They showed how the high contrast of the lines in the Lyα forest can arise naturally in structures resembling Zel'dovich pancakes (Zel'dovich 1970; Sunyaev & Zel'dovich 1972). The contrast arises from the overdensity of the gas, and the fact that the neutral density is proportional to the square of the total density.

In this paper, we use the Zel'dovich approximation and an *ad hoc* variant of it to calculate the probability distribution function of the Gunn-Peterson optical depth. This semi-analytical method can be useful for underdense regions where the gas has not yet been shocked in any collapse. In § 2 we describe our method, and in § 3 we present the results for various models of structure formation. § 4 shows how our results can be used to interpret the observations of the weakest lines in the Lyα forest as a fluctuating Gunn-Peterson effect, and discuss future applications of these observations as a probe to the primordial density fluctuations on small scales. Finally, in § 5 we summarize our results.

## 2. METHOD

In this section, we present the method by which we estimate the distribution of the optical depth to Lyα scattering caused by different regions in an inhomogeneous intergalactic medium, where the gas is falling into gravitational potential wells arising from initially gaussian fluctuations in the cold dark matter (CDM) distribution that grow by gravitational instability. We use the Zel'dovich approximation, and also what we call the "modified Zel'dovich approximation", by which we attempt to improve the prediction for the evolution of the density in underdense regions, or voids, as the rms fluctuation reaches values of order unity. In §2.1, the relation between the optical depth observed in a quasar spectrum and the density and velocity fields in the intergalactic medium is presented. § 2.2 shows the relation between the CDM and gas density fluctuations as affected by the finite gas pressure. In §2.3, we discuss the usual Zel'dovich approximation and our modification; and in §2.4, we describe the formalism to calculate the distribution of the density and the optical depth with these approximations, for gaussian initial density fluctuations.

### *2.1. The fluctuating Gunn-Peterson optical depth*
### *of an inhomogeneous intergalactic medium*

Throughout this paper, we assume that the gas in the intergalactic medium is photoionized and in ionization equilibrium with an ionizing radiation background of uniform intensity. Let the gas density at any point in space, with comoving coordinates $\mathbf{x}$, be $\rho_g(\mathbf{x}) = \bar{\rho}_g \, \nu_g(\mathbf{x})$, and let the density of neutral hydrogen be $\rho_n(\mathbf{x}) = \rho_{nu} \, \nu_n(\mathbf{x})$, where $\bar{\rho}_g$ is the average gas density in the universe, and $\rho_{nu}$ is the density of neutral hydrogen for a uniform gas of density $\bar{\rho}_g$ in ionization equilibrium with an ionizing background of the same intensity as in the nonuniform case. (For example, in an isothermal and mostly ionized medium, $\nu_n = \nu_g^2$.) The neutral hydrogen yields an optical depth due to Lyα scattering along the line-of-sight to a quasar which can be observed as a function of wavelength. The wavelength is related to the component of the total velocity of the gas along the line of sight, $V$, by $\lambda = \lambda_{\mathrm{Ly}\alpha}\,(1+z_0)\,(1+V/c)$, where $\lambda_{\mathrm{Ly}\alpha}$ is the Ly$\alpha$ wavelength, and the velocity obeys $V \ll c$, and is measured with respect to an arbitrary origin at redshift $z_0$.

For a uniform medium, this optical depth is given by (Gunn & Peterson 1965)

$$\tau_u = 4.65 \times 10^5\,\Omega_g\,h\,(\rho_{nu}/\bar{\rho}_g)\,(1+z)^3[H_0/H(z)]\;, \tag{1}$$

where $\Omega_g$ is the gas density divided by the critical density, $H_0 = 100\,h\,\mathrm{km\,s^{-1}\,Mpc^{-1}}$ is the Hubble constant at present, and $H(z)$ is the Hubble constant at redshift $z$. In the general case of a medium with varying density, the optical depth contributed by any point is increased proportionally to the neutral hydrogen density, and inversely proportionally to the gradient of the total velocity along the line of sight of observation. In general, there can be more than one point along the line of sight having a velocity $V$. Let these points have a comoving coordinate $x_{(\alpha)}$, where $x$ is the component along the line of sight. Ignoring the effects of thermal broadening, we then have

$$\tau(V) = \tau_u \sum_\alpha \nu_n(x_{(\alpha)})\,\frac{\dot{a}}{|dV/dx|_{x=x_{(\alpha)}}}\;, \tag{2}$$

where $a$ is the scale factor and dots denote derivatives with respect to cosmic time $t$, so that $H(z) = \dot{a}/a$. When the density fluctuations are small, there is a unique position to every velocity, but as the fluctuations grow velocity caustics will form, and within those $x(V)$ is multivalued. Velocity caustics generally appear earlier than caustics in real space (e.g., McGill 1990).

The derivative of the velocity appearing in equation (2) can be written as $dV/dx = (\partial V_i/\partial x_j)\,e_i\,e_j$, where $e_i$ is the unit vector along the line of sight. The total velocity is $\mathbf{V} = \dot{a}\mathbf{x} + \mathbf{v}$, where $\mathbf{v}$ is the peculiar velocity. Thus,

$$\frac{dV}{dx} = \left(\dot{a}\delta_{ij} + \frac{\partial v_i}{\partial x_j}\right)\,e_i\,e_j\;. \tag{3}$$

## 2.2. Growth of density perturbations in a gas with finite pressure

In order to encompass a number of the currently popular models for the growth of structure in the Universe, we allow for cold (CDM) and hot (HDM) dark matter components. The CDM has a negligible temperature and can grow perturbations on all scales of interest, which produce gravitational potential wells that affect the dynamics of the gas. The HDM, on the other hand, is so hot that its Jeans scale is much larger than the scales of interest, and it will be taken as uniform. In the present section, we can ignore the presence of the HDM.

The temperature of the gas in the intergalactic medium is sensitive to the history of reionization, since at low redshifts the thermal equilibrium time is longer than the Hubble time (e.g., Rees & Setti 1970). Assuming that the dominant heating and cooling mechanisms of the intergalactic gas are, respectively, photoionization heating and adiabatic cooling, Miralda-Escudé & Rees (1994) found that, for redshifts $z \lesssim 5$ and for the average baryon density, the gas temperature will tend to a value $T \approx T_0(1+z)^\beta$ within a Hubble time, where $T_0 \approx 3000\,\mathrm{K}$ and depends on the spectrum of the radiation, and $\beta \approx 0.9$. For other densities, the temperature is $T \propto \rho^{\beta/3}$ (see their Figs. 2 and 3). For mathematical convenience (see next paragraph), we take $\beta = 1$. Since the photoionization rate is $\propto \rho_n$, and the recombination rate is $\propto \rho_g^2 T^{-0.7}$ (assuming $\rho_n \ll \rho_g$), the equilibrium density of neutral gas obeys $\nu_n \approx \nu_g^{1.8}$.

In the linear regime (i.e., for small perturbations to a uniform-density background), and if the temperature decreases as $T = T_0(1 + z)$, then the power spectrum of the gas density fluctuations is related to that of the cold dark matter by

$$P_g(k) = \frac{P_{CDM}(k)}{[1 + (k/k_J)^2]^2},\tag{4}$$

with the Jeans wavenumber

$$k_J = \left( \frac{4\pi\mu m_p G \bar{\rho}_{CDM,0}}{\gamma k_B T_0} \right)^{1/2}.\tag{5}$$

Here, $\mu$ is the mean molecular weight in atomic units (= 0.6 for fully ionized hydrogen and helium in cosmological abundances), $m_p$ is the proton mass, $G$ is the constant of gravitation, $\bar{\rho}_{CDM,0}$ is the current average density of the cold dark matter, $\gamma$ is the value of the derivative $d\ln p/d\ln\rho$ in the perturbations ($\gamma = 4/3$ under the assumptions made above), and $k_B$ is Boltzmann's constant. A sketch of the derivation of this relation for a flat, CDM-dominated universe can be found in Peebles (1993, pp. 635-636), and the extension to the present, more general case is relatively straightforward. When $P_{CDM}(k)$ increases less rapidly than $k$ as $k \to \infty$ (which is expected to be the case because of smoothing processes before recombination), the velocity dispersion of the cold dark matter in collapsed objects decreases toward small scales; when this velocity dispersion drops below the sound speed of the gas, pressure prevents the gas from collapsing in the dark matter potential wells. The gas is therefore smooth on scales smaller than the Jeans mass, and the rms fractional fluctuations of the gas density have a finite value, which we define as $\sigma_J$.

In addition to the small-scale smoothing that appears already in linear perturbation theory, the pressure also affects the nonlinear evolution of the gas, and this can be important on scales not much larger than the Jeans scale. These effects cannot be included in our treatment.

### 2.3. Gravitational evolution

In this section, we study the gravitational evolution of the gas, treated as CDM with the power spectrum modified as in equation (4). Our treatment is for a general Friedmann-Robertson-Walker universe with arbitrary values of the density parameter $\Omega$ and the cosmological constant $\Lambda$. We take the CDM to have a fraction $f_{CDM}$ of the total mass, with the remainder being a uniform, static background (HDM).

It is useful to introduce the velocity variable $\mathbf{u} \equiv \mathbf{x}' = \mathbf{v}/(a\dot{a})$, where $\mathbf{x}$ are the comoving Eulerian coordinates, the primes denote derivatives with respect to the expansion factor $a$, and $\mathbf{v}$ is the peculiar velocity. The deformation tensor $\boldsymbol{\nabla}\mathbf{u} = \partial\mathbf{u}/\partial\mathbf{x}$ can be decomposed in the standard way,

$$\nabla_i u_j = \frac{1}{3}\theta\delta_{ij} + \Sigma_{ij} + \omega_{ij},\tag{6}$$

where $\theta = \boldsymbol{\nabla \cdot} \mathbf{u}$ is the trace, and the last two terms are, respectively, the traceless symmetric and antisymmetric parts of the tensor. The evolution of the normalized density $|\nu| = \rho/\bar{\rho}$ (the variable $\nu$ diverges and changes sign when a caustic is crossed) the velocity divergence $\theta$, the shear, and the vorticity are determined by

$$\nu' = -\theta\nu, \tag{7}$$

$$\theta' = -\frac{\theta^2}{3} - \frac{2-q}{a}\theta - \Sigma^2 + \omega^2 - \frac{3p}{a^2}f_{CDM}(\nu - 1), \tag{8}$$

$$\Sigma'_{ij} = -\frac{2-q}{a}\Sigma_{ij} - \frac{2}{3}\theta(\Sigma_{ij} + \omega_{ij}) - \Sigma_{ik}\Sigma_{kj} - \omega_{ik}\omega_{kj} + \frac{1}{3}(\Sigma^2 - \omega^2)\delta_{ij} - \frac{3p}{a}\left(\frac{\partial^2\varphi}{\partial x_i \partial x_j} - \frac{1}{3}\nabla^2\varphi\delta_{ij}\right), \tag{9}$$

and

$$\omega'_{ij} = -\frac{2-q}{a}\omega_{ij} - \Sigma_{ik}\omega_{kj} - \omega_{ik}\Sigma_{kj}. \tag{10}$$

Here, $\Sigma^2 \equiv \Sigma_{ij}\Sigma_{ij}$, $\omega^2 \equiv \omega_{ij}\omega_{ij}$,

$$p(a) = \frac{\Omega_0/2}{\Omega_0 + (1 - \Omega_0 - \Omega_{\Lambda,0})a + \Omega_{\Lambda,0}a^3}, \tag{11}$$

and

$$q(a) \equiv -\frac{a\ddot{a}}{\dot{a}^2} = \frac{\Omega_0/2 - \Omega_{\Lambda,0}a^3}{\Omega_0 + (1 - \Omega_0 - \Omega_{\Lambda,0})a + \Omega_{\Lambda,0}a^3} \tag{12}$$

is the cosmological deceleration parameter, given in terms of the present values (denoted, here and below, by a subscript 0) of the fractions of the closure density contributed by matter, $\Omega$, and by a cosmological constant, $\Omega_\Lambda = \Lambda/3H^2$ (e.g., Peebles 1993, p. 100). The gravitational potential $\varphi$ is the solution of the Poisson equation

$$\nabla^2\varphi = f_{CDM}(\nu - 1)/a. \tag{13}$$

The only term in equations (7) through (10) which does not only depend on the local values of the density and the spatial derivatives of the velocity field is the term proportional to $\partial^2\varphi/\partial x_i \partial x_j$ in eq. (9). This term makes it impossible to make an exact, local calculation of these quantities in terms of their initial values (Kofman & Pogosyan 1994). However, a few local approximations to these equations have been proposed and used (e.g., Zel'dovich 1970; Bertschinger & Jain 1994), with variable success. In this paper, we use the Zel'dovich approximation (Zel'dovich 1970; Shandarin & Zel'dovich 1989) and a variant of it, both of which are described below. All local approximations necessarily break down at orbit crossing.

The growing modes in linear theory have no vorticity, and if $\omega_{ij} = 0$ initially, it will be true until orbit crossing occurs (Peebles 1980, §22 and 23). Thus, we set $\omega_{ij} = 0$ in our approximations.

We write the Lagrangian coordinates of a given mass element as $\mathbf{y}$, defined so as to coincide with its comoving Eulerian coordinates $\mathbf{x}$ at the initial time, $a \approx 0$. Now, in both approximations we make the assumption that

$$\frac{\partial \mathbf{u}}{\partial \mathbf{y}}(\mathbf{y}, a) = D'(\mathbf{y}, a)\frac{\partial \mathbf{u}}{\partial \mathbf{y}}(\mathbf{y}, a = 0) \;, \tag{14}$$

i.e., that the time-evolution only multiplies the Lagrangian-space deformation tensor by a (possibly position-dependent) scalar $D'(\mathbf{y}, a)$. The Jacobian matrix of the transformation from Lagrangian to Eulerian coordinates is

$$\begin{aligned}\frac{\partial x_i}{\partial y_j}(\mathbf{y}, a) &= \delta_{ij} + \int_0^a \frac{\partial u_i}{\partial y_j}(\mathbf{y}, a')da' = \\ &\quad \delta_{ij} + D(\mathbf{y}, a)\frac{\partial u_i}{\partial y_j}(\mathbf{y}, 0) \equiv [1 - D(\mathbf{y}, a)\lambda_i]\,\delta_{ij} \;,\end{aligned} \tag{15}$$

where the last expression is valid in a local Cartesian coordinate system with axes aligned with the eigenvectors of the initial deformation tensor, and defines the eigenvalues $\lambda_i$. The determinant of this matrix has to be

$$\det\left|\frac{\partial x_i}{\partial y_j}\right| = (1 - D\lambda_1)(1 - D\lambda_2)(1 - D\lambda_3) = \nu^{-1} \;, \tag{16}$$

in order to satisfy the continuity equation (7). The Eulerian deformation tensor becomes

$$\frac{\partial u_i}{\partial x_j} = \frac{\partial u_i}{\partial y_k}\frac{\partial y_k}{\partial x_j} = -\frac{D'\lambda_i}{1 - D\lambda_i}\delta_{ij}. \tag{17}$$

The dynamical equation (eq. [8]) can be used to determine the evolution of $D$. Writing all variables in terms of $D$, and defining $\eta_1 \equiv \lambda_1 + \lambda_2 + \lambda_3$, $\eta_2 \equiv \lambda_1\lambda_2 + \lambda_1\lambda_3 + \lambda_2\lambda_3$, and $\eta_3 \equiv \lambda_1\lambda_2\lambda_3$, we have

$$D'' + \frac{2 - q}{a}D' = \frac{3f_{CDM}p}{a^2}\frac{\eta_1 D - \eta_2 D^2 + \eta_3 D^3}{\eta_1 - 2\eta_2 D + 3\eta_3 D^2} \;. \tag{18}$$

This equation, when linearized in $D$, becomes independent of the initial eigenvalues and reproduces the familiar *Zel'dovich approximation* (hereafter ZA; Zel'dovich 1970; Shandarin & Zel'dovich 1989). This is known to be the correct linear limit of the full Lagrangian dynamical equations (note that the problematic shear term does not contribute to this order), and gives the exact dynamics in the one-dimensional case. In this approximation, the full dynamics is described by a single growth factor $D(a)$, the same at all points of space. It gives a remarkably good description of the gravitational evolution of matter up to (but not beyond) shell crossing.

In our *modified Zel'dovich approximation* (MZA), we solve the nonlinear equation (18) in order to obtain $D(\boldsymbol{\lambda}, a)$ independently for each set of eigenvalues $\boldsymbol{\lambda} = (\lambda_1, \lambda_2, \lambda_3)$. This approximation gives the exact evolution as long as the velocity field and gravitational potential satisfy the tensor relation (obtained from evaluating eq. [9] in terms of $D$ and its derivatives, and comparing to eq. [18])

$$\nabla \mathbf{u} = \alpha(\mathbf{x}, a)\nabla\nabla\varphi, \tag{19}$$

where $\alpha(\mathbf{x}, a)$ is an arbitrary scalar function. This relation is satisfied by the growing modes in the linear regime (where both this approximation and the ZA reduce to the exact linear theory), but does not remain correct in the general case. However, it is always satisfied in some simple geometries, in particular for planar, cylindrical, and spherical symmetry. Thus, in these geometries the approximation gives exact results up to shell crossing. This makes it clearly better than the ZA, which only gives exact results in a planar symmetry. In particular, in a spherical void in an Einstein-de Sitter cosmology ($p = q = 1/2$) with $f_{CDM} = 1$, the ZA gives an asymptotic density decrease $\nu \propto a^{-3}$ instead of the correct result (reproduced by the MZA), $\nu \propto a^{-3/2}$. In fact, in a general void, if the tidal forces from collapsed objects near the edges of the void are neglected, particles should simply keep their original velocities in the limit where the density in the void is very low, and therefore the shape of the deformation tensor in Lagrangian coordinates should stay constant, which is the approximation (14). This suggests that, while the ZA generally exaggerates the emptiness of the voids, the MZA may describe them more accurately.

However, in addition to the inconvenient fact that the growth factor depends on the initial condition, the MZA has the problem that in cases when $\eta_1 < 0$ but not all three eigenvalues have the same sign, the integration will reach a singularity (corresponding to a zero of the denominator of the term on the right-hand side of eq. [18]) where $D'$ diverges, and beyond which the solution is not single-valued. Physically, the density initially decreases but reaches a minimum at which it turns around because the matter is contracting in at least one direction, finally leading to collapse in the ZA. In the MZA, the shear diverges at the turn-around point, leading to an unphysical singularity. In order to avoid this complication, we limit ourselves to cases where all $\lambda_i < 0$ (expansion in all directions) when using this approximation.

The MZA is similar to the approximation of Bertschinger & Jain (1994) in that it leaves the principal axes of the deformation tensor of a given Lagrangian matter element fixed in space. However, it differs in that the Bertschinger-Jain approximation predicts collapse into filamentary structures, whereas the MZA clearly produces "Zel'dovich pancakes."

Finally, we determine the initial conditions for equation (18). In the limit $a \to 0$, one has $D \propto a^\zeta$, with $\zeta = [(1 + 24f_{CDM})^{1/2} - 1]/4$ (Bond, Efstathiou, & Silk 1980). The value of the constant of proportionality in this relation is obtained as follows. We take $\langle \eta_1^2 \rangle = 1$, so the rms fractional density fluctuation of the gas at small scales (as calculated in linear perturbation theory) should be $\sigma_J = D(a)$ as obtained in the ZA. Thus, one can use the modified power spectrum from equation (4), together with the (directly or indirectly) measured amplitude at some other scale (e.g., the popular $\sigma_8$, the rms fractional density fluctuation in a sphere of radius $8h^{-1}$ Mpc, again in linear theory) to infer the present value of $D = \sigma_J$ in the ZA, fixing the normalization. In the MZA, we simply require the asymptotic behavior of $D(\mathbf{y}, a)$ as $a \to 0$ to be the same as that of $D(a)$ in the ZA.

*2.4. Probability distribution functions*

Doroshkevich (1970) calculated the PDF of the initial eigenvalues of the deformation tensor for a randomly chosen point in Lagrangian space for random Gaussian fields. We use this PDF to generate numerically a random set of values of the eigenvalues, for which we calculate the values of $\rho$ and $\tau$ expected in different models at different redshifts (as explained in §§2.1 - 2.3), and finally combine the obtained values to obtain PDFs for these two variables.

In order to generate random values for the eigenvalues $\lambda_i$, we find it convenient to change to new variables $R$, $S$, and $Z$ by

$$\lambda_1 = \frac{1}{3}\left\{ Z + \frac{R}{\sqrt{5}}\left[ S + \sqrt{3(1 - S^2)}\right]\right\},$$
$$\lambda_2 = \frac{1}{3}\left\{ Z - \frac{2RS}{\sqrt{5}}\right\}, \tag{20}$$
$$\lambda_3 = \frac{1}{3}\left\{ Z + \frac{R}{\sqrt{5}}\left[ S - \sqrt{3(1 - S^2)}\right]\right\}.$$

(Bertschinger & Jain 1994 use a similar change of variables.) In order to have $\lambda_1 \geq \lambda_2 \geq \lambda_3$, we require $0 \leq R < \infty$, $-1/2 \leq S \leq 1/2$, and $-\infty < Z < \infty$. Unlike the eigenvalues $\lambda_i$, these new variables are independent of each other, with joint probability distribution

$$\mathcal{Q}(R, S, Z) = \left(\frac{2}{3\sqrt{2\pi}}R^4 e^{-R^2/2}\right)\left(\frac{3}{2} - 6S^2\right)\left(\frac{1}{\sqrt{2\pi}}e^{-Z^2/2}\right), \tag{21}$$

which makes it easy to generate random values by standard algorithms (Press et al. 1992).

Kofman et al. (1994) point out that, in general, the density in Eulerian space is the result of adding the densities contributed by several particles ("streams") coming from different Lagrangian coordinates. This occurs after the first "pancakes" have collapsed. In the scenario of interest to us, the gas is shocked when that takes place, invalidating the approximations of §2.3 in these regions. The single-stream probability distribution of the eigenvalues in Eulerian space is $\mathcal{Q}(\lambda_1, \lambda_2, \lambda_3)/|\nu|$. Thus, the single-stream density PDF as sampled in Eulerian space takes the form

$$\mathcal{P}(\rho) = \int d\lambda_1\, d\lambda_2\, d\lambda_3\, \frac{\mathcal{Q}(\lambda_1, \lambda_2, \lambda_3)}{|\nu|}\delta\left(\rho - \bar{\rho}|\nu|\right) \tag{22}$$

(Kofman et al. 1994).

In order to obtain a PDF of the optical depth that can be compared with observations of quasar spectra, one must take into account that the latter do not sample the gas uniformly in Eulerian space, but rather in velocity, along the line of sight. It is convenient to rewrite eq. (3) as $dV/dx = \dot{a}\mu$, where (in Cartesian coordinates aligned with the principal axes of the gas element of interest)

$$\mu \equiv \left(\delta_{ij} + a\frac{\partial u_i}{\partial x_j}\right)e_i e_j = 1 - \frac{d\ln D}{d\ln a}\sum_i \frac{D\lambda_i e_i^2}{1 - D\lambda_i}. \tag{23}$$

The optical depth contributed by one mass element is

$$\tau = \frac{\tau_u \nu_n}{|\mu|}\ . \tag{24}$$

Now, we calculate the PDF of the optical depth in an "observational space" with velocity as the coordinate along the line of sight, and Eulerian distance in the transverse directions. In this space, the single-stream PDF of the eigenvalues for a fixed orientation of the line of sight with respect to the eigenvectors of the deformation tensor is $|\mu/\nu|\mathcal{Q}(\lambda_1, \lambda_2, \lambda_3)$, and the single-stream PDF of the optical depth can be written as

$$\mathcal{F}(\tau) = \int d\lambda_1 \, d\lambda_2 \, d\lambda_3 \, \frac{\sin\theta d\theta \, d\phi}{4\pi} \, \left|\frac{\mu}{\nu}\right| \, \mathcal{Q}(\lambda_1, \lambda_2, \lambda_3) \delta\left(\tau - \tau_u \frac{\nu_n}{|\mu|}\right) \ . \tag{25}$$

Here, $\theta$ and $\phi$ are angles giving the direction of the unit vector $e_i$. The single-stream PDFs are equal to the total (observable, multi-stream) PDFs, only if the contribution from regions which have gone through caustics in the respective spaces is negligible. In this case, it can be shown that the integral $N_s$ of the PDF over all possible parameter values is unity. In general, $N_s \geq 1$ is the average number of regions contributing to the optical depth at a given point in the spectrum.

As an estimate of the low-density end of $\mathcal{P}(\rho)$ and the low-optical depth end of $\mathcal{F}(\tau)$ in the general case, we use eqs. (22) and (25), except that we integrate only over those values of the integration variables which correspond to regions that have not yet gone through caustics in the space of interest, i.e., for which $1 - D\lambda_1 > 0$ in the case of $\mathcal{P}(\rho)$, and for which both $1 - D\lambda_1 > 0$ and $\mu > 0$ for $\mathcal{F}(\tau)$. For small $\rho$ and $\tau$, this should be a better approximation than integrating over the whole parameter space, since in the latter case we would also include regions which in our formalism have gone through a caustic and are reexpanding, but in reality are shocked as soon as they approach the caustic, and do not expand afterwards. Of course, the high-density regions are only shocked once they reach caustics in real (Eulerian) space rather than the "observational space" used to calculate $\mathcal{F}(\tau)$, but this should not make a difference for the behavior at small $\tau$.

## 3. RESULTS

We use the procedure described in §2 to obtain PDFs for four different models of structure formation: 1) standard CDM (sCDM) with $\Omega_0 = 1$ and $h = 0.5$; 2) mixed dark matter (MDM) with $\Omega_0 = 1$, $h = 0.5$, and $f_{CDM} = 0.7$; 3) CDM in an open universe (CDMo) with $\Omega_0 = 0.3$, $\Omega_{\Lambda 0} = 0$, and $h = 0.65$; and 4) CDM in a low density, flat universe (CDMΛ) with $\Omega_0 = 0.4$, $\Omega_{\Lambda 0} = 0.6$, and $h = 0.65$.

In order to normalize the growth factor in each of these models, we take the present value of the linearly extrapolated rms fractional mass fluctuation in a sphere of radius $8h^{-1}$ Mpc, $\sigma_8$, as inferred for each model from the fluctuations in the microwave background detected by COBE (see Bunn, Scott, & White 1994 for references). We use the values of $\sigma_8$ of Bunn et al. (1994) for sCDM ($\sigma_8 = 1.34$) and MDM ($\sigma_8 = 0.97$), that of Kamionkowski et al. (1994) for CDMo ($\sigma_8 = 0.35$), and that of Kofman, Gnedin, & Bahcall (1993) for CDMΛ ($\sigma_8 = 0.79$). It should be pointed out that Bunn et al. (1994) and Kamionkowski et al. (1994) use the full two-year COBE data, whereas Kofman et al. (1993) use the first-year data only. In addition, the fitting procedures of the latter two papers are not as accurate as those of the first.[3] However, these errors are comparable to the measurement errors and the effect of cosmic variance.

From $\sigma_8$ and the linear power spectrum, softened at small scales by pressure effects (eq. 4), we infer the value of $\sigma_J(a = 1) \equiv \sigma_{J0}$, the linearly extrapolated fractional gas mass fluctuation in a sphere of vanishing radius at the present time. We use the fit to the power spectrum of Bardeen et al. (1986) for sCDM, CDMo, and CDMΛ, and that of Klypin et al. (1993, note added in proof) for the cold component in an MDM model. The growth factor is then normalized to $D(a = 1) = \sigma_{J0}$ in the ZA (as discussed in §2.2, this also fixes the normalization in the MZA). Figure 1 shows $D(a)$ for all four models, assuming a temperature $T = 2000(1 + z)$ K. The models predict very different fluctuations on these small scales. Compared to these differences, the uncertainties due to the uncertain temperature are fairly small. For our four models, $D(a)$ decreases by $10 - 17\%$ if $T$ is increased by a factor of 3.

The density PDF in the ZA depends only on $D$, regardless of expansion parameter and cosmological model, and therefore the same PDF occurs at different times in different models. This is not exactly true for the PDF of $\tau$ in the ZA, which also depends (weakly) on $d \ln D/d \ln a$, because of the mapping from (Eulerian) physical space to velocity space. It is also not true for either PDF in the MZA, where the growth factor is different from point to point, and does not appear to be simply related to the global growth factor of the ZA. However, we have found that all these differences are quite small compared to the difference between models, and to the changes with redshift. Thus, the PDFs are fairly well characterized by a single number, the growth factor $D$ of the ZA.

Figure 2 illustrates the change in the density PDF with varying $D$. For each of the values, $D = 1/8$, $1/4$, $1/2$, and 1 (corresponding to the rms density fluctuation expected in linear theory), we show the full density PDF in the ZA (dashed curves), the density PDF in the ZA considering only fluid elements expanding along all three principal axes, i.e., for which all three $\lambda_i$ are negative (dotted curves), and the PDF in the MZA considering only fluid elements expanding along all three principal axes (solid curves). From the latter two sets of curves, it can be seen that, for progressively larger values of $D$, the fluid elements that expand along all three axes occupy a progressively larger fraction of the total volume. It is also clear from comparing the dashed and dotted curves that at low densities only the fluid elements that expand along all three principal axes are contributing.

The central result of this paper are the PDFs for $\rho$ and $\tau$ at different redshifts, shown in Figs. 3, 4, and 5, where the different types of curves have the same meaning as in Fig. 2. These figures illustrate that the typical density and optical depth in the interstellar medium are substantially lower than they would be if the gas were uniformly distributed. In addition, there are dramatic differences among models. Particularly low values are seen for the standard CDM model.

A comparison of the dashed and dotted curves in Figs. 3 – 5 also shows that the low-density (or low-optical-depth) tail of the PDFs comes nearly exclusively from fluid elements which expand along all three principal axes. These regions, although initially filling only $\sim 8\%$ of the volume, expand so much that they occupy most of it by $D \sim 2$. The similarity of the dashed and dotted curves almost up to the "knee" of the latter also suggests that it is acceptable to use only the fluid elements with $\lambda_1 \leq 0$ in the determination of the low density parts of the PDFs in the MZA. Of course, this does not guarantee the accuracy of the MZA itself, which should be tested by comparison with numerical simulations.

---

[3] After the present work had been submitted, Górski et al. (1994) reported a more accurate fit for open models. They obtain $\sigma_8 = 0.45 - 0.54$ for the particular model considered here, somewhat higher than the value we used. This makes the results for CDMo even more similar to those for CDMΛ; see below.

## 4. DISCUSSION

In the previous section, we have presented a calculation of the distribution function of the Gunn-Peterson optical depth, arising from underdense regions in a photoionized intergalactic medium where the gas is moving towards gravitationally collapsed objects.

The distribution of the optical depth can be directly determined from the observations of quasar spectra, in the region where the Ly$\alpha$ forest is observed. Some fraction of the spectrum is occupied by detectable absorption lines. These should correspond to overdense regions where the gas has collapsed and has been shocked. According to the simulations of Cen et al. (1994), the properties of the observed lines seem to be consistent with those arising from structures formed during gravitational collapse (sheets, filaments and halos) of the photoionized gas. Obviously, our analytical treatment cannot give us the distribution of optical depths from regions in the observed spectra occupied by such absorption lines.

In regions outside these absorption lines, one should observe the optical depth corresponding to the underdense gas that is left between collapsed structures, and has not been shocked. The distribution of optical depths in these regions, once the tails in the Voigt profiles of the strong lines have been subtracted, should correspond to the distribution we have calculated in this paper. However, we must bear in mind the limitations of our calculation when it is applied to the observations. First of all, our calculation of the optical depth caused by a given point of the intergalactic medium in Lagrangian space is only an analytical approximation. It is not clear how well our approximation of a constant ratio of the eigenvalues of the deformation tensor in underdense regions will work in practice. Second, only a fraction $f(\tau)$ of all the regions yielding a certain optical depth $\tau$ will not have collided with shocked gas, merging into systems yielding the absorption lines, and will also not be superposed with absorption from other regions having the same velocity along the line of sight. Here, we simply assume this fraction to be unity for low $\tau$, but in general $f(\tau)$ will be less than 1, and the fraction of the spectrum having optical depth lower than $\tau$ will be smaller than what we derive. Third, the hydrodynamic effects of the pressure have not been included, except for the smoothing of the power spectrum at the Jeans mass in the linear regime, which we have calculated under the assumption that $T \propto \rho^{1/3}$. The streaming of the gas away from the centers of voids should be slowed down by pressure gradients in the non-shocked gas; the importance of this effect was illustrated in the previous section by considering two different temperatures for the gas. Finally, thermal broadening will cause some smoothing of the Gunn-Peterson optical depth, and this will reduce fluctuations.

All of these effects imply that any results obtained from the observed intensity distributions using our analysis can only be taken as preliminary indications of the implications that may be derived from observations of the Gunn-Peterson effect. Our analytical results should be useful to better understand and interpret the numerical results that can be obtained with simulations like in Cen et al. (1994). The simulations are also subject to effects of finite resolution which may cause the underdensity of voids to be underestimated when the density gradients become too large. These simulations include all of the effects we have mentioned above, and they also predict the number of absorption lines at different column densities produced by collapsed regions. In order to match the number of predicted absorption lines in a specific theory with the observed ones, the quantity $\tau_u$ in equation (1) has to be fixed, and the amplitude of Gunn-Peterson fluctuations must then agree with observations.

Photoionized gas will first collapse in objects at the Jeans scale. The gas pressure prevents the collapse of smaller objects, causing a cutoff on the power spectrum that will cause a prominent structure of pancakes surrounding voids (Zel'dovich 1970). The separation between these pancakes should be of order the Jeans scale, $\lambda_J H(z) = 75 T_4^{1/2} \, \mathrm{km \, s^{-1}}$, where $T_4 = T/(10^4 \, \mathrm{K})$ is the temperature of the intergalactic medium, $H(z)$ is the Hubble constant at the epoch when the absorption lines are observed, and we have assumed the density in cold dark matter is equal to the critical value (e.g., Peebles 1980, eq. [16.5]). As larger scales continue to collapse, the voids become more empty and the sheets formed initially turn thinner as they accrete into more massive structures. However, because the gas distribution is smooth on the Jeans scale, there is always an intergalactic medium pervading all space between collapsed objects. The average separation between absorption lines corresponding to collapsed regions can never be smaller than $\lambda_J$, but can only increase if the sheets in voids become so thin that eventually their absorption lines are indistinguishable from the general Gunn-Peterson absorption.

The point we wish to emphasize here is that, as a consequence of the above paragraph, only the absorption lines that are, on average, separated by more than $\lambda_J$ can be associated with physically distinct structures, if the Ly$\alpha$ forest originates indeed from gravitational collapse. Whenever the abundance of the lowest column density lines rises above this value, these lines must correspond to the fluctuating Gunn-Peterson effect arising from a true intergalactic medium, rather than to individual clouds. If they were individual clouds, there is simply not enough mass for them to be above the Jeans mass given their abundance, and therefore they could only be pressure confined by an intergalactic medium much hotter than the gas in the clouds. One should notice that it is always possible to describe a continuum of Ly$\alpha$ absorption by a superposition of weak lines, as long as the column density distribution and the correlation function of such lines are left as free functions to be determined observationally. The fact that such fitting of superposed lines can be done does not prove that the lines are associated with real clouds.

It is interesting to notice that the Gunn-Peterson effect is the only method available to us to directly observe underdense matter in the universe. The usual detection of voids from the distribution of galaxies (e.g., Kirschner et al. 1981, 1987; Huchra et al. 1983) is done by noticing the absence of galaxies on a certain region, and these galaxies are overdense regions on a smaller scale than the voids being detected. In general, the primordial fluctuations on scales which have already entered the non-linear regime of gravitational collapse cannot be recovered from the distribution of matter in the overdense regions, because that is mostly sensitive to non-linear processes of gravitational collapse, and the memory of initial conditions is erased. However, the density of matter left in underdense regions depends on the initial fluctuations on small scales (e.g., Bernardeau 1992; see his Figure 10 for the evolution of the density probability distribution function). If the amplitude of fluctuations increases rapidly toward small scales, then the first objects should have collapsed at an earlier epoch, and there has been more time for the matter to flow out of voids. But if the power spectrum flattens toward small scales, so that the first objects collapsed more recently, then the median density in voids should be higher. Therefore, the determination of the distribution of the Gunn-Peterson optical depth, $\tau$, for low values of $\tau$, can be used to constrain the amplitude of primordial fluctuations on small scales, down to the Jeans scale at the temperature expected for photoionized gas in the intergalactic medium ($T \sim 10^4$ K).

The Gunn-Peterson absorption can be observed for two species: H I and He II. The He I effect, occurring at 584 Å, is expected to be very small and would be confused with hydrogen lines of lower redshift clouds; only a few lines corresponding to high column density systems have been observed (Reimers & Vogel 1993). We shall discuss the observations for the other two species.

### 4.1. The H I Gunn-Peterson effect

Most attempts to measure the Gunn-Peterson effect have been based on detecting the discrete Ly$\alpha$ lines, and then testing if the intensity observed between these lines is consistent with the extrapolated continuum of the quasar on the red side of the Ly$\alpha$ emission line (e.g., Steidel & Sargent 1987, Giallongo et al. 1994). Jenkins & Ostriker (1991) presented the distribution of intensities in several quasar spectra. They showed that in order to account for the total depression of the flux in the region of the Ly$\alpha$ forest, there has to be either a uniform Gunn-Peterson effect at the level of $\tau_{GP} \simeq 0.1$ at $z = 3$, or an equivalent absorption due to lines that are too weak and crowded to be detected. They also illustrated the effect caused by the presence of a uniform optical depth, $\tau$: there is a maximum value of the intensity that any pixel in the spectrum can reach, equal to $e^{-\tau}$ relative to the extrapolated continuum, except for the instrumental noise. If the absorption between the detected lines is to be explained instead by the superposition of weaker, undetected lines, then these lines cause not only an additional average depression, but also more fluctuations in the intensity, and some pixels are left with very little absorption (see Fig. 3 in Jenkins & Ostriker).

The higher resolution observations of Webb et al. (1992) of QSO 0000-263, at $z = 4.11$, clearly showed that no uniform Gunn-Peterson optical depth, at the level needed in the models of Jenkins & Ostriker where such a continuum absorption was postulated, can be present in the data. Webb et al. detected all the absorption lines with $N_{HI} > 10^{13.75}$ cm$^{-2}$, and looked at the intensity distribution in the regions outside these strong lines. They then modeled this distribution of intensities as arising from weaker lines, plus a possible uniform continuum. They found that a steep column density distribution, with $\beta = 1.7$ (where $f(N_{HI}) \, dN_{HI} \propto N_{HI}^{-\beta} \, dN_{HI}$) cannot be extrapolated to column densities $N_{HI} < 10^{13}$ cm$^{-2}$, because that would produce fluctuations of the intensity larger than observed. However, if the shallower slope of $\beta = 1.3$ is assumed for lines with $N_{HI} < 10^{13.75}$ cm$^{-2}$, and is extrapolated to zero column density, then both the average depression and the intensity fluctuations can be explained, without the need of any uniform absorption. If a continuum absorption is present, then the number of weak lines must be even lower, since they must produce a smaller average depression, but there must be more of them at $N_{HI} \gtrsim 10^{13}$ cm$^{-2}$ in order to obtain the observed intensity fluctuations.

This turnover in the distribution of column densities in the Ly$\alpha$ forest, required by the data of Webb et al. (1992), should be expected in a model of gravitational collapse, arising from the lower limit to the size of gravitationally collapsed objects, given by the Jeans length (this turnover is seen in Figure 3 of Cen et al. 1994 at $N_{HI} \simeq 10^{12.5}$ cm$^{-2}$, although it is probably too pronounced in this case because of the method used to select the lines). In fact, absorption lines with $N_{HI} \gtrsim 10^{13}\,\mathrm{cm}^{-2}$ at $z = 4$ have an average abundance of about one line every $100\,\mathrm{km\,s}^{-1}$ in the observed Ly$\alpha$ forest, as is easily found from the column density distribution of the observed lines, and the evolution with redshift (see, e.g., Petitjean et al. 1993; Murdoch et al. 1986; Lu, Wolfe, & Turnshek 1991; Jenkins & Ostriker 1991; Bechtold 1994). This is of the order of the expected Jeans length and, as discussed above, these lines must roughly correspond to the weakest ones that can arise from structures that have gravitationally collapsed in at least one dimension. Lines with even lower column density should form a continuum of absorption, *and this true continuum would not be separated into individual lines if the resolution and the signal-to-noise were perfect.*

We then conclude that if intergalactic space at $z \sim 4$ is not filled by gas at $T \gg 10^4\,\mathrm{K}$, heated by processes other than photoionization and confining the Ly$\alpha$ clouds, then *a significant part of the fluctuations and the median decrement of the intensity distribution in Figure 2 of Webb et al. (1992) must arise from the fluctuating Gunn-Peterson effect.* Moreover, the upper limit of $\tau_{GP} < 0.06$ in Webb et al. is not applicable, because their central assumption of the uniformity of the Gunn-Peterson absorption is *not* valid.

Following our line of argument, we make the following simple assumption: all the absorption lines with $N_{HI} < 10^{13}\,\mathrm{cm}^{-2}$ in the models used by Webb et al. to fit their intensity distribution correspond to the fluctuating Gunn-Peterson effect. The distribution of optical depths that these lines produce can then be directly compared to our calculations presented in § 3. The first model of Webb et al. assumes $\beta = 1.7$, and a sudden cutoff at $N_{HI} = 10^{13}\,\mathrm{cm}^{-2}$. A uniform Gunn-Peterson effect of $\tau_{GP} = 0.04$ is then needed. In our interpretation of their results, this represents a *lower* limit to the average optical depth, since in other models incorporating a lower uniform continuum the number of lines with $N_{HI} > 10^{13}\,\mathrm{cm}^{-2}$ needs to be reduced, and therefore a larger fraction of the total decrement in the intensity must come from weaker lines, which we consider as part of the Gunn-Peterson effect.

Thus, in the second model in Webb et al. , if no lower cutoff is assumed for the column density distribution, they find that a value of $\beta = 1.3$ for $N_{HI} < 10^{13.75}\,\mathrm{cm}^{-2}$ fits their distribution. The average optical depth produced by all these lines is about 0.25 (see Figure 2 of Webb et al. ), and approximately 30% of this optical depth is due to lines with $N_{HI} < 10^{13}\,\mathrm{cm}^{-2}$. We then derive the higher value $\tau_{GP} = 0.08$.

The turnover of the column density distribution should probably occur smoothly, over a wide range in column density, and the optical depth in the Gunn-Peterson effect should then be intermediate between the two models discussed above. We therefore adopt a value of $\tau_{GP} = 0.06$ at $z = 4$, and assume that this has to be equal to the median optical depth calculated in § 3 for different models.

We start considering our open CDM model, and the CDM model with a cosmological constant. Both these models predict the linearly extrapolated rms density fluctuation on the Jeans scale, $\sigma_J$, to be about 1 to 1.5 at $z = 4$. From the solid lines in Figures 4 and 5, we see that the Gunn-Peterson optical depth is $\tau \lesssim 0.1\tau_u$ over 25% of the spectrum. If we extrapolate the solid line to higher $\tau$, assuming it would have a similar shape as the dashed line for the Zel'dovich approximation if regions with positive eigenvalues of the deformation tensor were included, 50% of the spectrum should have $\tau \lesssim 0.2\tau_u$. This should correspond to the observed median Gunn-Peterson optical depth, so we derive $\tau_u \simeq 0.3$ at $z = 4$ for these models. On the other hand, for the standard CDM model, the fluctuations on the Jeans scale are $\sim 3$, and the median value of the Gunn-Peterson optical depth should be only $\tau \sim 0.01\tau_u$, implying the much larger value $\tau_u \simeq 6$. For models with lower amplitude of fluctuations at the Jeans scale, like the MDM model, the required value of $\tau_u$ is lower.

In terms of the baryon density of the universe $\Omega_b$, and the intensity of the ionizing background, $J_{HI,-21}$ (defined as in eq. [4] of Miralda-Escudé & Ostriker 1992), the optical depth $\tau_u$ in equation (1) would be given as:

$$\tau_u = 2.3 \left(\frac{\Omega_b h^2}{0.015}\right)^2 \frac{0.5}{h} \left(\frac{1+z}{5}\right)^6 \frac{H_0}{H(z)} \frac{T_4^{-0.7}}{J_{HI,-21}} , \qquad (26)$$

where $T_4$ is the gas temperature in units of $10^4$ K, $H(z)$ is the Hubble constant at redshift $z$, and we have assumed a helium abundance $Y = 0.24$, and that the helium is doubly ionized. For the open CDM model we use here, $H(z)/H_0 = 7.4$ at $z = 4$, and for the model with a cosmological constant, $H(z)/H_0 = 7.1$. Fixing $\Omega_b$ to the value required by primordial nucleosynthesis, we need $J_{HI,-21} \simeq 1$ for these two models, in good agreement with determinations of this intensity from the proximity effect (Bajtlik, Duncan & Ostriker 1988; Lu, Wolfe, & Turnshek 1991; Bechtold 1994). On the other hand, for the standard CDM model, we need $J_{HI,-21} \simeq 0.2$.

The simulations of Cen et al. (1994) were done for the same CDM model with cosmological constant that we have used here. According to their results, the number of absorption lines is correctly predicted if $J_{HI,-21} \simeq 1$, which is the same value that we find is needed to explain the Gunn-Peterson effect. Thus, this model seems to be satisfactory, and probably any models with rms density fluctuations on the Jeans scale near unity should give similar results.

If the density fluctuations are larger, like in standard CDM with the COBE normalization, then systems on larger scales will have collapsed, which will have higher total column densities of gas. The column densities of neutral hydrogen would be even more increased relative to models with smaller density fluctuations, because a lower intensity of the ionizing background is required to explain the amplitude of the Gunn-Peterson effect. Thus, the turnover in the column density distribution would occur at a column density that is too high. Such models would therefore be in conflict with the observations. Their basic difficulty is that there is too much gas in collapsed systems on large scales, and too little left in the intergalactic medium, so that the contrast between the high column density lines and the Gunn-Peterson effect is too large.

At the opposite extreme, if the density fluctuations are too small, there should be too few collapsed systems producing high column density absorption lines, and the Gunn-Peterson effect would be too large, or equivalently, there would be too many superposed weak lines, giving a very steep column density distribution. Models of this type, such as the MDM model, also have difficulty in forming a sufficiently large number of halos where the gas can dissipate and form damped Ly$\alpha$ systems and high redshift quasars with the observed numbers (Mo & Miralda-Escudé 1994; Kauffmann & Charlot 1994; Ma & Bertschinger 1994).

We then conclude that models with an rms density fluctuation of order unity at the Jeans scale are favored by the observations of the Ly$\alpha$ forest, and by the distribution of intensities in Webb et al. (1992). As we have discussed before, this is only a preliminary result, which needs to be further investigated with numerical simulations. However, it seems clear that the ratio of the Gunn-Peterson optical depth to the number of individual absorption lines will constrain the primordial rms density fluctuations as we have discussed in this paper.

### 4.2. The He II Gunn-Peterson effect

For the He II Gunn-Peterson effect, all our equations are equally applicable as long as most of the helium in the intergalactic medium is twice ionized. In this case, the density of singly ionized helium is also approximately proportional to $\nu^{1.8}$, as discussed in § 2, since the recombination coefficient for He II depends on temperature almost in the same way as the H I recombination coefficient. The abundance of He I is always very small, and has a negligible effect on the equilibrium of He II. Equation (1) is changed to

$$\tau_{u, HeII} = 2.5 \times 10^3 \, \frac{\Omega_b h^2}{0.015} \, \frac{0.5}{h} \, \left( \frac{1+z}{4} \right)^3 \, \frac{H_0}{H(z)} \, y_{HeII} \; . \tag{27}$$

Jakobsen et al. (1994) have recently reported the first observation of the spectrum of a quasar at a rest-frame wavelength immediately below 304 Å. A cutoff in the flux was observed at this wavelength, and interpreted as the He II Gunn-Peterson effect. The trough is observed at $z \sim 3$, and the flux decrement is more than a factor of 5 at the two-sigma level.

From our discussion above, it is clear that in a theory of gravitational collapse for the Ly$\alpha$ forest, such a trough must be due to the Gunn-Peterson effect, since any weak lines causing it would have to be overlapping. However, because of the fluctuations in the Gunn-Peterson effect, we expect that some regions in the spectrum should have low optical depth, and show up as "gaps" in the He II trough, which could be observed in a spectrum of better resolution and signal-to-noise than in Jakobsen et al. (1994). Such gaps should be a very sensitive probe to the most underdense voids in the intergalactic medium.

In general, the lower limit required on the fraction of singly ionized helium in the intergalactic medium to give a certain He II Gunn-Peterson optical depth is larger than in the case of a uniform medium, because the density of the gas is much lower than average in the voids. However, the constraints on the spectrum of the ionizing background are not modified, since this depends only on the ratio of the H I and He II Gunn-Peterson effects.

## 5. CONCLUSIONS

In this paper, we have considered the model of gravitational collapse from primordial density fluctuations for the Ly$\alpha$ forest, where photoionized gas in the intergalactic medium collapses into the structure forming on scales above the Jeans mass. The observations of the Gunn-Peterson effect and the low column density lines in the Ly$\alpha$ forest are unique among all other cosmological observations, in the sense that they allow us to directly observe matter in underdense regions, as well as moderately overdense matter in the infall regions around collapsed objects. An analytical approximation was used to calculate the distribution of the density field and the Gunn-Peterson optical depth from regions that are initially expanding in all directions in comoving coordinates.

We have shown that, in this theory, the distribution of intensities observed by Webb et al. (1992) in the $z = 4.1$ quasar QSO 0000-263 gives evidence that the fluctuating HI Gunn-Peterson effect has been detected. Over most of the spectrum of a quasar outside strong lines the optical depth originates from underdense regions, and its typical value, which is found to be $\tau_{GP} \simeq 0.06$ at $z = 4$, is therefore significantly lower than the value $\tau_u$ expected for a uniform intergalactic medium with the baryon density implied by primordial nucleosynthesis.

Models for structure formation in which the fractional rms density fluctuation is close to unity at the Jeans scale at $z = 4$ imply a value of $\tau_u$ that is consistent with the measured intensity of the ionizing background from the proximity effect, $J_{HI} \simeq 10^{-21} \, \mathrm{erg \, s^{-1} \, cm^{-2} \, Hz^{-1} \, sr^{-1}}$. From the results of Cen et al. (1994), these models are also consistent with the observed column density distribution of the absorption lines, for the same value of $J_{HI}$. A much higher or much lower value for the density fluctuations should lead to different predictions for the relative strength of the fluctuating Gunn-Peterson effect and the individual absorption lines. These conclusions are at this point preliminary, but further studies with numerical simulations for several models, and better observations of high redshift quasar spectra at high resolution, should give more definite conclusions.

There are three complications that must be considered, which can affect the properties of the intergalactic medium in ways that cannot be incorporated in numerical simulations using only basic physical principles. The first is the fact that the intensity and the spectrum of the ionizing background will fluctuate due to random variations in the number of emitting sources, as well as the number of absorbers (Zuo 1992a,b; Fardal & Shull 1993). The second is that the temperature of the intergalactic medium should also fluctuate, and may not be a function of the density only, because it depends on the initial temperature to which the gas is heated at reionization. This initial temperature can vary substantially from place to place, depending on the spectrum of the sources reionizing the medium (Miralda-Escudé & Rees 1994). In addition, the spectrum of the ionizing background should also vary depending on the random fluctuations in the number of sources and absorbers, and this affects the equilibrium temperature of the gas. Finally, energetic events in the collapsed objects, such as supernova explosions and quasars, can eject gas into the intergalactic medium and heat it to high temperatures.

The latter process is the one that can change more drastically the properties of the Gunn-Peterson effect. If ejection of gas is important, then the voids can be refilled and more absorption lines can be produced by clouds not associated with gravitational collapse. Such lines could constitute a different population of objects, and should have smaller transverse sizes than the gravitationally collapsing systems. The Gunn-Peterson effect could decrease much more slowly with time if gas were constantly ejected into voids. Recent observations of CIV in systems with $N_{HI} \simeq 3 \times 10^{14} \, \mathrm{cm^{-2}}$ (Cowie et al. 1995; Fan & Tytler 1995) suggest that the metallicity is not very low compared to systems of higher column density. If this is also true for much lower column density lines in the Ly$\alpha$ forest, then the intergalactic medium must have been contaminated with metal-rich, ejected gas at some point in the past. This could have occurred at a very early epoch, when the intergalactic medium was being reionized. In that case, the metals could have been ejected from systems formed from neutral gas on very small scales, and the ejection velocities might be too small to significantly affect the evolution of the voids in the intergalactic medium. However, if the metal enrichment was from systems that can eject gas at velocities comparable to those generated by the collapsing structures, then the assumptions we have made here would be invalidated. This could then allow models with high initial density fluctuations to be made consistent with observations of the Gunn-Peterson effect.

If it is found that the simple case where all these complicated effects do not have a strong influence on the intergalactic gas can predict the observed properties of the Ly$\alpha$ forest, then it should be clear that most of the volume in the intergalactic medium at $z = 4$ should not have been affected by shock waves from explosions. Fluctuations in the gas temperature and the intensity of the background cannot drastically alter the median Gunn-Peterson optical depth. The amplitude of the Gunn-Peterson fluctuations should then provide a strong constraint on the primordial fluctuations at small scales.

We wish to thank Martin Rees and David Weinberg for many illuminating and pedagogical discussions, and Francis Bernardeau, Karl Fisher, Roman Juszkiewicz, Marc Kamionkowski, Jeremiah Ostriker, Douglas Scott, Eli Waxman, and the referee, Xavier Barcons, for interesting comments and conversations. A. R. is supported by NSF (grant PHY 92-45317) and by the Ambrose Monell Foundation, and J. M. by NASA grant NAG-51618 and by the W. M. Keck Foundation.

# REFERENCES

Bajtlik, S., Duncan, R. C., & Ostriker, J. P., 1988, ApJ, 327, 570

Bardeen, J. M., Bond, J. R., Kaiser, N., & Szalay, A. S. 1986, ApJ, 304, 15

Bechtold, J. 1994, ApJS, 91, 1

Bernardeau, F. 1992, ApJ, 392, 1

Bertschinger, E., & Jain, B. 1994, ApJ, 431, 486

Bond, J. R., Efstathiou, G. P. E., & Silk, J. 1980, Phys. Rev. Lett., 45, 1980

Bond, J. R., Szalay, A. S., & Silk, J., 1988, ApJ, 324, 627

Bunn, E. F., Scott, D., & White, M. 1994, ApJL, in press

Cen, R. Y., Miralda-Escudé, J., Ostriker, J. P., & Rauch, M. 1994, ApJ, 437, L9

Cowie, L. L., Songaila, A., Kim, T., & Hu, E. M. 1995, submitted to AJ

Doroshkevich, A. G. 1970, Astrofizika, 6, 581

Fan, X. M., & Tytler, D. 1995, preprint

Fardal, M. A., & Shull, J. M. 1993, ApJ, 415, 524

Giallongo, E., et al. 1994, ApJ, 425, L1

Giallongo, E., Cristiani, S., Fontana, A., & Trevese, D. 1993, ApJ, 416, 137

Górski, K. M., Ratra, B., Sugiyama, N., & Banday, A. J. 1994, ApJL, submitted

Gunn, J. E., & Peterson, B. A. 1965, ApJ, 142, 1633

Huchra, J. P., Davis, M., Latham, D. W., & Tonry, J. 1983, ApJS, 52, 89

Ikeuchi, S. 1986, Astrop. Space Sci., 118, 509

Jakobsen, P., Boksenberg, A., Deharveng, J. M., Greenfield, P., Jedrzejewski, R., & Paresce, F. 1994, Nature, 370, 35

Jenkins, E. B., & Ostriker, J. P. 1991, ApJ, 376, 33

Kamionkowski, M., Ratra, B., Spergel, D. N., & Sugiyama, N. 1994, ApJ, 434, L1

Kauffmann, G., & Charlot, S. 1994, ApJ, 430, L97

Kirshner, R. P., Oemler, A., Schechter, P. L., & Shectman, S. A. 1981, ApJ, 248, L57

———— 1987, ApJ, 314, 493

Klypin, A., Holtzman, J., Primack, J., & Regös, E. 1993, ApJ, 416, 1

Kofman, L., Gnedin, N., & Bahcall, N. 1993, ApJ, 413, 1

Kofman, L., & Pogosyan, D. 1994, ApJ, submitted

Kofman, L., Bertschinger, E., Gelb, J. M., Nusser, A., & Dekel, A. 1994, ApJ, 420, 44

Lu, L., Wolfe, A. M., & Turnshek, D. A. 1991, ApJ, 367, 19

Ma, C. P., & Bertschinger, E. 1994, ApJ, 434, L5

McGill, C. 1990, MNRAS, 242, 544

Miralda-Escudé, J., & Ostriker, J. P. 1992, ApJ, 392, 15

Miralda-Escudé, J., & Rees, M. 1994, MNRAS, 266, 343

Mo, H. J., & Miralda-Escudé, J., 1994, ApJ, 430, L25

Murdoch, H. S., Hunstead, R. W., Pettini, M., & Blades, J. C. 1986, ApJ, 309, 19

Peebles, P. J. E., 1980, *The Large-Scale Structure of the Universe* (Princeton Univ. Press: Princeton)

Peebles, P. J. E., 1993, *Physical Cosmology*, (Princeton Univ. Press: Princeton)

Petitjean, P, Webb, J. K., Rauch, M., Caswell, R. F., & Lanzetta, K. 1993, MNRAS, 262, 499

Press, W. H., Teukolsky, S. A., Vetterling, W. T., & Flannery, B. P. 1992, Numerical Recipes in Fortran (2d ed.; Cambridge: Cambridge Univ. Press)

Reimers, D., & Vogel, S. 1993, A&A, 276, L13

Rees, M. J. 1986, MNRAS, 218, 25p

Schneider, D. P., Schmidt, M., & Gunn, J. E. 1991, AJ, 101, 2004

Shandarin, S. F., & Zel'dovich, Ya. B. 1989, Rev. Mod. Phys., 61, 185

Steidel, C. C., & Sargent, W. L. W. 1987, ApJ, 318, L11

Sunyaev, R. A., & Zel'dovich, Ya. B. 1972, A&A, 20, 189

Webb, J. K., Barcons, X., Carswell, R. F., & Parnell, H. C. 1992, MNRAS, 255, 319

Zel'dovich, Ya. B. 1970, A&A, 5, 84

Zuo, L. 1992a, MNRAS, 258, 36

Zuo, L. 1992b, MNRAS, 258, 45

# FIGURE CAPTIONS

**Fig. 1:** Fractional rms density fluctuations at the Jeans scale, $\sigma_J$, calculated using linear perturbation theory, as a function of the expansion factor, $a = (1+z)^{-1}$, in the four models of structure formation considered in this paper. The assumed baryon temperature is $2000(1+z)$ K ($\sigma_J$ also corresponds to the growth factor $D(a)$ in the Zel'dovich approximation).

**Fig. 2:** Cumulative probability distribution function (PDF) of the density (i.e., fraction of the volume at densities lower than a certain value $\rho$) as a function of $\log(\rho/\bar{\rho})$, where $\bar{\rho}$ is the average density, for different values of the fractional rms density fluctuation calculated in linear theory, $\sigma = 1/8, 1/4, 1/2$, and 1. The dashed lines were obtained in the Zel'dovich approximation (ZA), considering all matter elements. The dotted lines also correspond to the ZA, but considering only those matter elements which expand along all three principal axes. The solid lines were obtained in our *modified Zel'dovich approximation* (MZA) using a flat, CDM-dominated cosmological model, and again considering only matter elements which expand along all three principal axes.

**Fig. 3:** Cumulative PDFs of the density for four models of structure formation described in the text. The meaning of the line types is the same as in Fig. 2. Within each type, the four lines represent (from right to left) redshifts $z = 4, 2, 1$, and 0. The assumed gas temperature is $T = 2000(1+z)$ K.

**Fig. 4:** Cumulative PDFs of the optical depth (i.e., fraction of a quasar spectrum sampling matter of an optical depth lower than a given value $\tau$), as a function of $\log(\tau/\tau_u)$, where $\tau_u$ is the optical depth expected if the matter were uniformly distributed and expanding with the Hubble flow. The models and the redshifts are the same as in Fig. 3 (but note that the lower redshift lines are off the scale in some of the plots). Line types are as in Fig. 2. The assumed gas temperature is $T = 2000(1+z)$ K.

**Fig. 5:** The same as Fig. 4, except at a gas temperature $T = 6000(1+z)$ K, which increases the Jeans scale by $3^{1/2} \approx 1.7$, and therefore lowers the amplitude of the fluctuations.